# Analysing the Effect of Recommendation Algorithms on the Amplification of Misinformation

Miriam Fernández[a,1], Alejandro Bellogín[b], Iván Cantador[b]

[a]*Knowledge Media Institute, The Open University, United Kingdom*
[b]*Escuela Politécnica Superior, Universidad Autónoma de Madrid, Spain*

**Abstract**

Recommendation algorithms have been pointed out as one of the major culprits of misinformation spreading in the digital sphere.[2] However, it is still unclear how these algorithms really propagate misinformation, e.g., it has not been shown which particular recommendation approaches are more prone to suggest misinforming items, or which internal parameters of the algorithms could be influencing more on their misinformation propagation capacity. Motivated by this fact, in this paper we present an analysis of the effect of some of the most popular recommendation algorithms on the spread of misinformation in Twitter. A set of guidelines on how to adapt these algorithms is provided based on such analysis and a comprehensive review of the research literature. A dataset is also generated and released to the scientific community to stimulate discussions on the future design and development of recommendation algorithms to counter misinformation. The dataset includes editorially labelled news items and claims regarding their misinformation nature.

## 1. Introduction

Misinformation [3] is affecting our democracies, polarising our societies, manipulating our economies, and putting in danger our health and environment

---

[1] Corresponding author.

[2] Please note that amplifying, spreading and propagating are indistinctly used in this paper to refer to the amplifying effect that recommendation algorithms may have on misinformation when recommending or suggesting items to users.

[3] In this paper we use the term *misinformation* to refer to misleading information, hoaxes, conspiracy theories, hyper-partisan content, click-bate headlines, pseudoscience and false news.*1*

[1].[4] Today, in the context of a global pandemic, misinformation has led to tragic results, including links to assaults, arson and deaths.[5] It is a common problem in all media, but it is exacerbated in digital social media due to the speed and ease in which posts are propagated [2]. The social web enables users, bots and malicious actors to spread information rapidly without confirmation of truth, and to paraphrase this information to fit their intentions and beliefs [3].

Online misinformation is a high-dimensional socio-technical problem with multiple influencing factors, including: (i) the ways in which **information** is constructed and presented [4, 5], (ii) the **users**' personality, values, emotions and susceptibility [6, 7] as well as the presence of bots and malicious accounts [8, 9], (iii) the architectural characteristics of the **digital platforms** where such information is propagated (i.e., the structure of the social networks, constraints on the type of messages and sharing permissions, etc.) [10], and (iv) the **algorithms** that power the recommendation of information within those platforms [11].

Recommendation algorithms (RAs) have been heavily criticised for filtering the information observed by users, who may be placed into biased filter bubbles where the only content they access is the type of content they like and is generated by other people with similar opinions [12].[6] This comes as a consequence of the fact that RAs are part of the so-called *feedback loop*, i.e., systems that aim to reinforce a cycle that attempts to optimise user retention and interactions.

Additionally, these algorithms tend to rely on engagement signals for the recommendation of information (such as user preferences on topics, social connections between users, and the relatedness between the topics in social networks [13]), and are therefore affected by popularity and homogeneity biases [14, 15]. In this context, filter bubbles and biases may limit the exposure of users to diverse points of view [16] and reduce the quality of the information the users access [17], potentially making them vulnerable to misinformation. Youtube, for example, has been criticised for amplifying videos that are divisive, sensational and conspiratorial.[7]

---

[4] YouTube's algorithm is pushing climate misinformation videos, and their creators are profiting from it: https://www.niemanlab.org/2020/01/youtubes-algorithm-ispushing-climate-misinformation-videos-and-their-creators-are-profiting-from-it/, Engagement with Deceptive Outlets Higher on Facebook Today Than Run-up to 2016 Election https://www.gmfus.org/blog/2020/10/12/new-study-digital-new-deal-findsengagement-deceptive-outlets-higher-facebook-today Last Accessed: Jan 2021

[5] Coronavirus: The human cost of virus misinformation: https://www.bbc.co.uk/news/ stories-52731624 Last Accessed: Jan 2021

[6] Up Next: A Better Recommendation System: https://www.wired.com/story/creatingethical-recommendation-engines/ Last Accessed: Jan 2021

[7] Fiction is outperforming reality': how YouTube's algorithm distorts truth https://www.

Despite these criticisms, there is an important gap in the research literature when it comes to understanding the impact that RAs have on the spread of false and misleading information [18, 11]. Some works have studied the effect that RAs may have on the creation of filter bubbles [12, 19, 20], and others have created models to understand the effect that common popularity biases in RAs may have on the quality of items consumed by users [17]. However, a more in-depth investigation is needed to better understand which of these algorithms are more prone or susceptible of spreading misinformation, under which circumstances, and how the internal functioning of such algorithms could be modified or adapted to counter their misinformation recommendation behaviour. This is a very complex issue, and previous attempts have resulted in harmful effects. For example, Twitter modified its RA to recommend popular tweets into the feeds of people who did not subscribe to the accounts that posted those tweets. This change, which provides popular opposing views, was heavily criticised for amplifying inflammatory political rhetoric and misinformation.[7]

Misinformation is thus a problem with a high number of dimensions that interrelate to one another [2, 21], some of them affecting what RAs learn and therefore, how they will behave. For this reason, adapting RAs to counter their misinformation spreading behaviour requires an in-depth understanding not only of the internal mechanisms of such algorithms, but also of the data they manipulate, the users they serve, and the platforms they operate in.

Starting from this position, this paper investigates the effect of some of the most popular recommendation algorithms on the spread of misinformation on Twitter. A set of guidelines on how to adapt these algorithms is provided based on the performed analysis and a deep review of the research literature. In our investigation, a dataset is created and released to the scientific community to stimulate discussions on the future design and development of RAs to counter misinformation.[8] The dataset includes editorially labelled news items and claims regarding their misinformation nature.

Our contributions can be summarised as follows:

- An exhaustive analysis of previous work studying (i) the dimensions of the misinformation ecosystem that may affect the performance, results and biases of RAs, (ii) the role of RAs on the spread of misinformation, and (iii) effective strategies to counter misinformation and correct misperceptions.

---

theguardian.com/technology/2018/feb/02/how-youtubes-algorithm-distorts-truth Last Accessed: Jan 2021

[7]How Twitter's algorithm is amplifying extreme political rhetoric https://edition.cnn.com/2019/03/22/tech/twitter-algorithm-political-rhetoric/index.html Last Accessed: Jan 2021

[8]The URL to the dataset and code will be released upon acceptance.

- The creation of a dataset containing Twitter user profiles, items, ratings and misinformation labels that enables studying the effect of RAs on the amplification of misinformation.

- A comprehensive analysis of different state-of-the-art RAs frequently used in industry and academia (including collaborative filtering techniques like nearest neighbours and matrix factorisation) on the amplification of misinformation, by means of three evaluation metrics proposed to account for such amplification: misinformation count, ratio difference, and Gini.

- A set of guidelines on how to modify and adapt RAs based on the conducted analysis as well as on the review of the research literature.

We note that the focus of this work is on recommendation algorithms and not on recommender systems. The latter include other aspects aside from the algorithm (e.g., user interface) that are not considered in our study. The remainder of the paper is structured as follows. Section 2 discusses related work on the field. Section 3 discusses a series of challenges derived from this problem, and our proposed approach to address them. Section 4 describes the dataset that we have generated for experimentation and that we are making available to the research community. Section 5 presents the different RAs that have been assessed, as well as the metrics and methods used to assess them. Sections 6 and 7 present our results and recommendations on how to adapt RAs to palliate the misinformation amplification effect. Discussions and conclusions are presented in Sections 8 and 9 respectively.

**2. Related Work**

Misinformation is a multifaceted (human, sociological and technological) problem, and it has been focus of investigation in several research fields including social sciences, journalism, computer science, psychology and education. In this section, we aim to provide a summary of the literature based on four dimensions of the misinformation problem related with the design and building of RAs: *content*, *users*, *platform characteristics*, and *algorithms*. We complement this analysis with a summary of some of the strategies that have been found effective in correcting misperceptions for a more informed analysis on the adaptation of RAs.

Although our contributions are more in line with those works that have attempted to understand the impact of algorithms on the spread of misinformation (see Section 2.1.4) a multi-dimensional review of the literature considering other dimensions like content, users, and platforms, is needed for the design of comprehensive RAs adaptation guidelines (see Section 7). We note that ours does not aim to be an exhaustive literature review on misinformation. For a comprehensive overview of the problem, the reader is referred to the following surveys: [22, 23, 24, 25, 26, 27].

*2.1. Misinformation Dimensions*

*2.1.1. Content*

Content is an important factor of the misinformation problem, and also a key aspect to consider in the design, evaluation and adaptation of RAs. Items to be recommended can be present in various *forms* (as news articles, research papers, blog entries, and social media posts) and discuss a wide range of *topics*, such as health, elections and disasters, to name a few. These items are not only textual, but sometimes include information in different *formats*, like images or videos. Moreover, combinations of these formats are frequently used to propagate misinformation (e.g., a news title linked with an image from a different place, or from a different time). The *framing* of misinforming news also varies between false news, rumours, conspiracy theories, and misleading content [5]. Other important elements to consider about content are their *emotional tone*, their *origin* (news outlets, social contacts, public figures, etc.) as well as the *time* when they are posted. Note that recency is particularly relevant to the recommendation of news items.

Multiple works in the literature have attempted to understand the characteristic of such content, and to develop algorithms that could automatically detect it. Castillo and colleagues [28, 29, 30] studied information credibility on Twitter mainly based on content features, and created supervised machine learning classifiers to detect credibility. Their studies concluded that credible tweets tend to include more URLs, and are longer than non-credible tweets. Question and exclamation marks tend to concentrate on non-credible tweets, frequently using first and third-person pronouns. Qazvinian and colleagues [31] also studied content features for misinformation detection. They concluded that lexical and Part of Speech (POS) patterns are key for correctly identifying rumours. Hashtags can result in high precision but lead to low recall. Hansen [32] showed how messages news with negative sentiment became more viral. In terms of topics, Vosoughi and colleagues [7] showed how false political news have a more pronounce cascading effect than false news about terrorism, natural disasters, science, urban legends, or financial information.

Initiatives driven by the journalism research field have also attempted to identify key features of misinforming content. Credibility Coalition[8] published an article in 2018 [33] listing credibility indicators for news articles. These include content indicators, such as the use of clickbait titles, or the use of emotionally charged tone, and context indicators, such as the representation of sources cited in the article. Extensions and selections of these signals have also be captured by the W3C Credible Web Community Group.[9]

---

[8] https://credibilitycoalition.org

These studies have derived on the creation of tools that attempt to identify misinforming content automatically. Examples include: TweetCred,[10] ClaimBuster,[11] or the Global Disinformation Index.[12] These tools rely on the characterisation of misleading content, and on manually compiled lists of misleading articles and websites, such as the one generated by Zimdars [34]. Various media literacy projects have also emerged from these studies to teach users how to identify misleading content.[13] Games have also been developed for this purpose. Fakey,[14] for example, simulates a typical social media news feed and asks the users to recognise suspicious content. More recently, the University of Cambridge has also released Go Viral[15] as a way to help users to recognise false COVID-19 related information.

*2.1.2. Users*

Users are a key dimension of the misinformation problem, and a core aspect of the functioning of RAs. Researchers have therefore studied the effect of different *motivations* [35], *personalities* [36], *values* [37] and *emotions* [7], and their effect on misinformation, as well as the *susceptibility* of users to spread misinformation [38, 39] and to interact with malicious actors [40]. For example, extroverts and individuals with high cooperativeness and high reward dependence are found more prone to share misinformation [35]. Psychology also shows that individuals with higher anxiety levels are more likely to spread misinformation [41]. The attention that users pay also plays an important role. Studies have shown how information overload and limited attention seem to contribute to the spread of misinformation [42].

Another important aspect to consider when it comes to users is the appearance of malicious actors within an information ecosystem, such as bots and sock puppets accounts. Social bots play a disproportionate role in spreading articles from low credibility sources. They also target users with many followers through replies and mentions, manipulating them to reshare misinformation [8, 9]. Users are also sometimes hired to support and propagate arguments or claims simulating grassroots social movements [43]. This phenomenon is known as crowdturfing. One of the latest examples is the campaign uncovered by The

---

[9] https://www.w3.org/community/credibility/
[10] https://chrome.google.com/webstore/detail/tweetcred/fbokljinlogeihdnkikeeneiankdgikg?hl=en
[11] https://idir.uta.edu/claimbuster/
[12] https://disinformationindex.org/the-index/
[13] https://ec.europa.eu/digital-single-market/en/news/winners-european-media-literacy-awards
[14] https://fakey.iuni.iu.edu/
[15] https://www.cam.ac.uk/stories/goviral

Washington Post, which was enlisting teens to spam a pro-Trump social media agenda for the 2020 US elections.[16]

As with content detectors, several tools have emerged in the last years that aim to identify malicious actors within social networks automatically. Examples include Botometer,[17] which checks the activity of a Twitter account, and gives a score indicating how likely is for the account to be a bot, and BotSlayer,[18] a tool that supports journalists and relevant stakeholders to discover coordinated campaigns in Twitter. Other tools like misinfo.me [44] encourage users to self-reflect by providing them with an assessment of how they have been interacting with misinformation, and which accounts among the users they follow are spreading more misinformation.

*2.1.3. Platforms and Social Networks*

Platforms that distribute online information are designed differently and, therefore, facilitate the spread of misinformation in different ways. *Content limitations* (e.g., Twitter and its 280 character maximum length for posts), *ability to share information* and select the subsets of users with whom such information is shared (*sharing permissions*), the ability to *vote* (e.g., Reddit) or to *express emotions* for content (e.g. Facebook), are important aspects of platform design that may shape the content, the way information spreads, and the *social network structure*. The typology and topology of the social network are indeed key factors of misinformation dynamics [23] and also important in the design of RAs. Note that RAs take into consideration not only similarity between items but also between users, as well as social connections, when generating recommendations.

In this context, researchers have focused on understanding how misinformation flows across different social networks. Shao and colleagues [9], for example, analysed the spread of 400 thousand articles on Twitter for ten months in 2016 and 2017. They concluded that low-credibility sources spread through original tweets and retweets, while few are shared in replies (i.e., the spreading patterns of low-credibility content are less conversational). They also observed that some accounts in the network acted as "super-spreaders" posting a low credibility article hundreds or even thousands of times, suggesting that the spread is amplified through automated means. Vosoughi [7] analysed a dataset of rumour cascades on Twitter and concluded that false news diffused significantly farther, faster, deeper and more broadly than the true ones. Kwon and colleagues [45] highlighted the fact that rumours have fluctuations over time, and that there are key structural differences in the spread of rumours vs. non-rumours. In addition to these studies, automatic tools like Hoaxy,[19] have also emerged to help monitoring the spread of misinformation.

---

[16] https://www.washingtonpost.com/politics/turning-point-teens-disinformation-trump/2020/09/15/c84091ae-f20a-11ea-b796-2dd09962649c_story.html
[17] https://botometer.osome.iu.edu/
[18] https://osome.iu.edu/tools/botslayer/

*2.1.4. Algorithms*

The full details of the algorithms that social networking sites have developed to personalise and recommend information to users are not known to the public. Their primary goal is, however, to increase user engagement and time spent on the platform, as a way of maximising revenue from ads shown in news feeds. Economic interest behind advertising ecosystems, and their effect on misinformation, have been at the core of recent studies [46], documentaries,[20] and campaigns.[21]

Critics of the RAs behind social networks [16] have emphasised the fact that users do not decide what they see, but are exposed only to the information that those algorithms select from them, introducing users in so-called filter bubbles. Since recommendation algorithms are designed to provide us with information that we like, based on our past interactions, and based on people who are similar to us, we risk ending up in bubbles where we only receive information that is pleasant, familiar and confirms our beliefs. We may not see the diverse set of opinions and information potentially available in the network. Additionally, since past interests determine what we are exposed to in the future, this may be leaving less room for the unexpected encounters that spark creativity, innovation, and the democratic exchange of ideas. Furthermore, users may not even be aware of this information filtering process. A 2015 study conducted with 40 Facebook users indicated that 62% of those users were entirely unaware of any curation, believing instead that every single story from their friends and followed pages appeared in their news feed [47].

RAs are known to suffer from popularity bias (i.e., the algorithm promotes information that is trending on the platform - e.g., getting more clicks) [14, 15]. In addition to the potential effect of item popularity, the information that users consume in social networking sites is also influenced by two other types of biases: (i) social biases and (ii) cognitive biases, particularly confirmation bias. The first one refers to the fact that the information that users are exposed to mainly comes from friends or accounts that the users follow. The second one refers to the fact that users are more likely to consume information that agrees with their own beliefs.

Several works in the literature have reported empirical studies and network simulations to understand whether filter bubbles do indeed exist in social media and whether these are the effect of RAs. In a 2015 study, Nikolov and colleagues [12] confirmed the presence of social bubbles on Twitter. They showed that collectively, people access information from a significantly narrower spectrum of sources through Twitter compared to a search baseline. A similar study [19] showed how Facebook's three filters (the social network, the algorithm, and a user's own content selection) decrease exposure to ideologically challenging news when compared to a random baseline. The article concludes that the composition of the users' social network is the most important factor affecting the mix of content encountered on social media with individual choice also playing a large role. The news feed (i.e., the algorithm effect) has a smaller impact on the diversity of information according to this study.[22] A recent 2020 study confirms this effect claiming that, under the presence of homophily (i.e., users preferring interactions and social ties with individuals that are similar to them), echo chambers and fragmentation are system-immanent phenomena [20].

---

[19] https://hoaxy.iuni.iu.edu/
[20] TheSocialDilemma:https://www.netflix.com/gb/title/81254224
[21] https://www.slpnggiants.com/

Further works have studied the effect of algorithmic popularity bias on the quality of information that users consume. Based on a designed model, Ciampaglia and colleagues [17] concluded that popularity bias hinders average quality when users are capable of exploring many items, as well as when they only consider very few top items due to scarce attention.

These works have aimed at understanding the effect that RAs may have on the creation of filter bubbles [12, 19, 20], as well as the effect that common popularity biases may have on the quality of items consumed by users [17]. The studies show how filter bubbles and popularity biases may make users more vulnerable to misinformation by reducing the diversity and quality of the information they are exposed to. *Our work aims to advance the state of the art by analysing how different RAs may influence the spread of misinformation, and under which conditions. We do not account here for social or cognitive biases, just algorithmic effects. Our hypothesis is that by better understanding these algorithms, and how they behave under the presence of misinformation, we can propose more informed adaptations to counteract the effect of false and misleading content.*

*2.2. Strategies to Correct Misperceptions*

Adaptations of RAs can be broadly focused on: (i) reducing the number of misinforming items they recommend and, (ii) adapting them to recommend information that could potentially help correcting misperceptions. Understanding successful and unsuccessful strategies to counter misperceptions is therefore key to propose more informed adaptations of RAs. We next survey studies that have attempted to correct misperceptions.

In the past decade, more than 110 independent fact-checking groups and organisations emerged online around the world [48] (e.g., Full Fact in the UK, Snopes and Root Claim in the US, FactCheckNI in Northern Ireland, and Pagella Politica in Italy, to name just a few). These groups and organisations aim to provide a frontline service in dealing with false information online and providing appropriate corrections.

---

[22] https://research.fb.com/blog/2015/05/exposure-to-diverse-information-onfacebook-2/

Existing research indicates that presenting people with corrective information is however likely to fail in changing their salient beliefs and opinions, or may, even, reinforce them [49, 50]. People often struggle to change their beliefs even after finding out that the information they already accepted is incorrect or misleading. Human beings strive for internal psychological consistency. We tend to favour information that confirms and supports our previous beliefs and values (confirmation bias). Inconsistency, on the other hand, tends to become psychologically uncomfortable (cognitive dissonance) and we tend to reject it [51].

Nevertheless, some strategies have been found to be effective in correcting misperceptions [52], such as exposing users to related but disconfirming stories [53], or revealing the demographic similarity of the opposing group [49]. In the context of health misinformation, Vraga and Bode [54] argue that "observational correction" is an accurate strategy at changing misconceptions. Observational correction refers to the fact that those who witness a correction on social media, but are not directly engaged with the misinformation item, are less affected by cognitive dissonance, and thus more amenable to correction. Vraga and Bode also suggested: (i) citing highly credible factual information with links to expert sources, (ii) offering a coherent alternative explanation for the misinformation, (iii) using multiple corrections to reinforce the message, and (iv) trying to correct misinformation early, before misperceptions are entrenched. In the context of more polarised topics, such as political misinformation, it is however unclear whether corrections work, or even worsen the problem for users who are unwilling to revise their believes [50, 55].

We have considered these suggestions when proposing adaptations to RAs (see Section 7).

## 3. Proposed Analysis Approach

Our primary goal in this paper is the analysis of the impact that different RAs have on the amplification of misinformation. To the best of our knowledge, no previous work has attempted to target this problem, which has required to confront a wide range of challenges, from dataset building, to user and item modelling, and algorithm and metric selection. In this section, we briefly describe some of these challenges, and in subsequent sections, we will detail how we have approached such challenges, and which design decisions we have considered.

**Data Collection:** Generating datasets that enable studying how different RAs amplify misinformation constitutes a significant challenge. These datasets should contain information about users, items, ratings (i.e., user-item interactions) and labels about which of those items are misinformation. To the best of our knowledge, existing datasets in the literature either: (i) provide a set of labelled misinforming items (e.g., datasets generated by fact-checker organisations) - without providing information about users or user-item interactions (ratings) or,

(ii) provide social media data collections (e.g., Twitter datasets, which contain information about users and items) but they either do not provide labels about which of those items are misinformation, or do not provide information about user-item interactions (ratings). Details of some of these publicly existing datasets and an analysis of their limitations, as well and the design decisions and methodology followed to build the dataset generated in this work, are provided in Section 4.

**Modelling Users and Items:** When studying RAs it is particularly important to define user and item profiles [56]. User profiles can be captured by considering either explicit or implicit feedback. Explicit feedback may consist of ratings or topics of interest stated by the user. Approaches based on implicit feedback to model a user's profile, in contrast, generally capture information from user interactions with the system (browsing sessions, clicks, etc). As we have seen in Section 2.1.2, within the misinformation problem, it may also be worth exploring how to capture multiple additional elements as part of user profiles including: the users' susceptibility, their emotions, personalities, motivations, values, or even information about their similarity with potential malicious actors (bots, sock-puppets). In terms of item modelling, multiple representations could also be considered: using the full content of the news items, a summary, or even some tags or categories assigned to the items. As we analysed in Section 2.1.1, within the misinformation problem, other important aspects could also be captured as part of the item profiles, including: form, format, topics, framing, emotional tone, origin and time. More information about how users and items have been modelled in this study is given in Section 5.

**Algorithm Selection:** A key aspect of our research is to understand which recommendation algorithms are more prone to suggest misinformative items to users. To investigate this, we needed to select a representative pool of recommendation techniques to test against appropriate datasets, among the well-known collaborative filtering (CF), content-based (CB) and hybrid approaches [57]. The recommender systems literature is mostly focused on CF, since it can be applied to any domain, only requiring user-item interactions, not needing additional item features or metadata. Studying these methods may help us to better understand whether the user-item interactions could help, by themselves, to spread or avoid misinformative items, since these models neglect any information about the items and their misinformative features. Hence, for the purpose of this study, we have decided to concentrate on this first set of methods. Here, it is important to highlight that, to the best of our knowledge, the impact of recommendation algorithms on the amplification of misinformation has not previously been studied. Hence we first need to understand how some of the most commonly used algorithms could behave in these scenarios (see Section 5.2). The algorithms selected for this study are, in fact, the building blocks of real-world recommender systems in industry and academia [58].

**Metrics:** We needed to be able to assess how different recommendations where increasing/decreasing the amount of misinformation provided to the users. Section 5.3 contains a description of the metrics we have proposed for that purpose.

The following sections describe the process followed to generate the dataset for our experiments, as well as the performed analysis set-up and the obtained results.

**4. Dataset Creation**

As highlighted earlier, generating datasets that enable to study how different RAs amplify misinformation constitutes a significant challenge. These datasets should contain data about items (i.e., available news or pieces of information), users, user-item interactions (or ratings) indicating which users have interacted with which pieces of information, and labels expressing which of those pieces of information are indeed misinformation. An illustration of these components is presented in Figure 1. Note that, for the purpose of this work we are considering explicit user-interactions, such as posting or sharing data, as ratings.

In the next subsections we present: (i) the investigation conducted to identify potential existing datasets, (ii) the methodology used to construct the dataset presented in this work and, (iii) the strengths and limitations of the generated resource.

*4.1. Analysing Existing Datasets*

We conducted a comprehensive investigation of datasets that could contain the above mentioned components, particularly focusing on datasets containing labelled data, which is generally very expensive to obtain. Various initiatives in the literature, such as Media Futures[23], have also attempted to gathered sim-ilar datasets, particularly focused on recent COVID-19 misinformation. A list of these datasets is available in their website. [24] While containing labelled information about misinforming items, the datasets do not provide comprehensive data about users and user-item interactions.

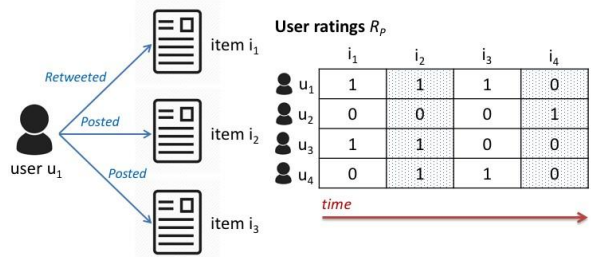

Figure 1: Modelling user profiles. User-ratings indicate which items the user has interacted with (i.e., posted or shared). A user profile contains 1 for item $i_x$ if the user has interacted with the item and 0 otherwise. The grey colour in the matrix (items $i_2$ and $i_4$) indicate that such items are misinformation.

---

[23] https://mediafutures.eu/

Some datasets provide automatically generated labelled data (see the related work sections of [59, 60] for a list of these datasets). However, automatically labelled data may introduce a high degree of noise. Therefore, we decided to concentrate on datasets that have been manually labelled, preferable by factcheckers.

In our investigation we also reviewed datasets from the multimedia evaluation community, like NewsReel[25] or FakeNews.[26] While these datasets contain items (news), and user-item interactions, they do not contain labelled information about such items. Additionally, the Fake News dataset is also only available as part of the Fake News detection competition.

What we observed after this analysis is that existing datasets in the literature either: (i) provide a set of labelled misinforming items (e.g., datasets generated by fact-checker organisations) – without providing information about users or user-item interactions (ratings) or, (ii) provide social media data collections (e.g., Twitter datasets, which contain information about users and items) but they either do not provide labels about which of those items are misinformation, or do not provide comprehensive information about user-item interactions (ratings). In some cases, a handful of interactions per user are provided (e.g., the COAID dataset[28]) by gathering the names of the users who posted known misinforming items. However, the timelines (user-interactions) of these users are not collected and analysed to investigate additional interactions of the same users with other misinforming items.

A dataset that is worth discussing here is Fakenewsnet [61].[27] This dataset contains labelled data (fake news) gathered from fact checkers, such as Polifact[28], as well as the tweets (items) that contain such fake news. The authors of those tweets (users) can be recollected via the Twitter API, as well as their most recent timelines (ratings). However, the misinforming items were identified and collected back in 2018. This means that, while the dataset contains users and ratings, there is a mismatch between the ratings that can be obtained (which are recent), and the labelled data, which covers news published around 2018 or earlier. It is therefore not possible to know which of the recent items the users have interacted with, are indeed misinformation. In recent days (26th of January 2021) Twitter announced that existing restrictions to data collection via its API will be removed for academic research, allowing the full gathering of historical data.[29] Our plan is therefore revisiting this dataset for future research.

---

[24] https://mediafutureseu.github.io/datasets.html
[25] http://www.multimediaeval.org/mediaeval2018/newsreelmm/
[26] https://multimediaeval.github.io/editions/2020/tasks/fakenews/
[28] https://github.com/cuilimeng/CoAID
[27] https://github.com/KaiDMML/FakeNewsNet
[28] https://www.politifact.com/

After this analysis, we selected a set of datasets that, although incomplete for the purpose of this study, could help us developing our own dataset. Table 1 presents a list of the datasets that were selected and reused, as well as their key characteristics. As we can see, while all these datasets contain labelled data, none of them provides rating information. The methodology followed to generate our final dataset is presented in the next section.

*4.2. Dataset Design and Creation Methodology*

As mentioned in the previous section, to the best of our knowledge, there are currently no available datasets in the literature to investigate the impact of RAs on the amplification of misinformation. The US National Science Foundation, the University of Colorado, the University of Minessota, and Northwestern University, have recently been collaborating and gathering use cases from the research community to develop the necessary datasets and research infrastructure to enable the investigation of new problems on the field of News Recommenders.[30] While we, as many other researchers, have input into this initiative, the datasets and infrastructure are still not available.

Hence, for the purpose of this work, and as a first step to progress on the investigation of how different RAs may impact the spread of misinformation,

Table 1: Publicly available datasets.

| Dataset | Description | Users | Items | Ratings | Labels |
|---|---|---|---|---|---|
| NELA-GT-2018 [62] | 713 articles collected between 02/2018-11/2018 directly from 194 news and media outlets including mainstream, hyper-partisan, and conspiracy sources | No | 713 (textual articles – no URLs) | No | Yes |
| CoronaVirusFacts Alliance dataset[33] | COVID-19 falsehoods generated by 88 fact-checkers in 74 countries since Jan 24th, including articles published in at least 40 languages | No | 8,624 (URLs + text + social posts) | No | Yes |
| Misinfo.me [44] | ClaimReview annotations from different fact-check publishers and aggregators | No | 34,764 (URLs + text) | No | Yes |
| Covid-19 myths [59] | two Tweets referring to two fails claims (holding your breath can diagnose COVID-19 and gargles with warm water, salt or vinegar can wash away COVID-19) | No | 1,493 tweets | No | Yes |
| CMU-MisCov19 [60] | 4,573 manually annotated tweets over 17 themes around the COVID-19 discourse | 3,629 | 4,573 tweets | No | Yes |

we have generated our own dataset.

---

[29] https://blog.twitter.com/developer/en_us/topics/tools/2021/enabling-the-future-of-academic-research-with-the-twitter-api.html
[30] https://docs.google.com/forms/d/e/1FAIpQLSf3h7-syPSUvsp13BV6ULUb2EMCpUT5io_NzaMD5Sbhb0Fwmg/viewform

The design decisions and methodology followed to create this dataset are explained below:

**Social Networking Platform:** We selected Twitter as a social networking platform to conduct this study because it enables, via its APIs, to collect users, items and ratings (in this case the 3,200 most recent items that each user has either posted or shared). Note that other platforms (such as Facebook or Whatsapp) only allow collecting data from public groups or pages, which means that user-item interactions (ratings) would be limited and incomplete.

**Labels (misinforming claims)**: As we can observe by the descriptions of the datasets (see Table 1), misinforming claims can be in the form of tweets, URLs, or text. This means that *multiple tweets, URLs, or texts can match the same false claim* (e.g., *"holding your breath for 10 seconds can diagnose COVID19"*). For the purpose of our work, we merged the claims of the CoronaVirusFacts Alliance dataset, Misinfo.me, Covid-19 two myths, and CMU-MisCov19 datasets. To do so, we first selected only claims labelled as false. Note that these datasets also contain labels for claims that are "unknown" or fact checked as "true." After filtering, we merged those datasets. To not duplicate information, we took into consideration: (i) if the URLs were the same, (ii) if the tweet IDs were the same, (iii) if the tweet ID and the parent ID were the same (so that retweets were matched with their original tweet), and (iv) if the texts were the same (exact matching) or did have a similarity score greater than 0.9 using Sentence-BERT [63]. We ended up with a list of 59,525 false claims, which are expressed as 42,876 URLs, 63,043 pieces of text, and 1,214 tweet IDs.

**Items:** It is important to consider that each tweet is not a different item. An item is a piece of information (news). This means that tweets containing the same information (either in the form of text, URL, or retweet) are considered as the same item.

**Users and user ratings:** As illustrated in Figure 1, we needed to collect a series of users with their corresponding user-item interactions (or ratings). We considered that if the user had posted or retweeted (shared) an item, such user had a preference (rating=1) for that item. To select a reliable list of Twitter users, we started with the CoronaVirusFacts Alliance dataset. This dataset has been fully generated by fact-checker organisations considering data from 74 countries, and it is our most reliable starting data point. We selected all the tweets labelled as false in this dataset (476), and extracted the authors of those tweets. We then took a random sample of users who had retweeted those tweets, increasing the dataset to 6,884 users. Note that via the Twitter search API it is only possible to collect 8 days back. The selected users were therefore biased to those that retweeted such content after 1st September 2020 (the CoronaVirusFacts Alliance dataset was made available to research collaborators on the 8th of September 2020). The idea of doing a random sample (restricted to a maximum of 20 retweets per tweet) was to avoid gathering users within the same social neighbourhood (i.e., users following other users), since our aim here is to test the algorithmic effect, not the potential effect of social or cognitive biases (see Section 2.1.4). Once we had the IDs of the 6,884 users, we proceed to collect their timelines using the Twitter API (the maximum number of tweets that can be collected per user was 3,200).

Within the collected tweets, we then proceed to search for the 59,525 false claims previously gathered. All users had 15 or less misinforming items in their timelines. To reduce the sparsity of the dataset, we selected items in English and (from the items that were not labelled as misinformation) those containing URLs. Our decision to filter English items was to facilitate exploiting any kind of RA – such as content-based – to be tested against the dataset. The decision of filtering items containing URLs was based on the assumption that items containing URLs are more likely to point out to information (news). The final dataset contains 2,921 users, 1,014,004 items, and 1,116,658 interactions (ratings) – see Table 2.

*4.3. Strengths and Limitations of the Generated Dataset*

As we have discussed in the previous sections, generating datasets that enable to study how different RAs propagate misinformation poses significant challenges. Misinforming items need to be compiled across multiple fact-checker organisations. The same item (e.g. "bleach cures COVID-19") can appear in the form of social media posts, URLs, or texts. These different forms of the same item need to be identified and grouped. Users, spreading such items in social media, in this case Twitter, need to be found, and their timelines have to be collected. Many careful considerations have been put into the generation of our dataset. However, it is also important to highlight that the generated dataset also presents several limitations.

Users have been selected based on the Coronavirus Facts Alliance Dataset. Although this dataset covers misinformation (identified by fact-checkers) from 74 countries, it is specific to COVID-19. Users who have never spread COVID19 misinformation are not covered in our data. Users in our dataset have very low numbers of misinforming items in their timeline, which may not be entirely representative of a social network. One may expect to see some users spreading a lot of misinformation, while many may spread none [64]. The creation of synthetic datasets could be a potential solution for future work, even though real effects would need to be measured via user studies, which opens up new challenges in terms of privacy and ethics. Finally, the dataset is focused on Twitter, since gathering user timelines (i.e., user ratings) from other platforms is limited. For instance, Facebook or Whatsapp only allow collecting data from public groups and pages, which means that only few messages (user interactions) per user can be collected.

Despite these limitations, it is important to stress that assessing the impact of different Recommendation Algorithms (RAs) on the propagation of misinformation is a problem not previously targeted despite its significance. Existing datasets and baselines are, to the best of our knowledge, nonexistent. The developed dataset is the first resource of its kind, capturing user profiles, ratings and misinformation labels, and enabling conducting offline evaluations of RAs and their impact on misinformation propagation.

More information on how the dataset has been used to conduct our study is presented in the following sections.

## 5. Evaluation Setup

In this section, we describe how we have exploited the dataset described in the previous section to simulate possible scenarios under different proportions of misinformative items in the system and/or shared by each user (Section 5.1). Besides, we describe the evaluated recommendation algorithms (Section 5.2), and explain how we propose to account for the presence of misinformation in the recommendations (Section 5.3). To foster reproducibility, we will make the code and dataset publicly available.

### 5.1. User Profiles

As described before, the first step when collecting information to build the dataset was to identify users who originally shared (i.e., *tweeted*) claims that were explicitly labelled as misinformative items. Because of the scarcity of tweets labelled by fact-checkers (only 476 tweets within the CoronaVirusFacts

**Algorithm 1:** Ratio-based user profile generator

**Function** *generate user u, ratio r* neg ← { i ∈ u: i is misinformative } ;
  // Negative claims
  neu ← u \ neg ;                              // Neutral claims
  desNeg ← r · |u| ; //Desired negative ratio desNeu ← (1 - r) · |u| ;
   //Desired neutral ratio
  **while** *(desNeg > |neg|) OR (desNeu > |neu|)* **do**
    **if** *desNeg > |neg| ; //Downsampling negative* **then** desNeg ← desNeg - 1;
    **end**
    **if** *desNeu > |neu| ; // Downsampling neutral* **then** desNeu ← desNeu - 1;
    **end** newTotal ← desNeg + desNeu; desNeg ← r · newTotal;
    desNeu ← (1 - r) · newTotal;

  userProfile(u) ← sample(neg, desNeg) ∪ sample(neu, desNeu);
  **end**
**end**

Alliance), the limitations of the used APIs (mainly restrictions on the temporal availability of the messages), and the reduced set of matches between the collected false claims and tweets, our dataset may not be entirely capturing the reality of the social network. Note that all users have between 1 and 15 ratings associated with misinformative items. This is a situation that may not be entirely representative [64]. However, to have some control on the amount of information exploited by the algorithms, we included the following constraints. First, we imposed a ratio $r$ that every user should satisfy regarding the amount of misinformative vs. neutral (either non-informative or unknown) items; for instance, $r = 0.5$ means that every user should have as many misinformative as neutral items. Second, we allowed sub-sampling to match the desired level of misinformation ratio, both in terms of misinformative or neutral items. An algorithm to achieve this is presented in Algorithm 1, where its main idea is to obtain the maximum number of either types of items that satisfy a given ratio.

Finally, to control against the base scenario where no constraints are imposed, we considered the special value $r = \emptyset$ as the situation where no filter is applied, that is, all the users in the dataset are transformed into user profiles and considered for training the recommendation algorithms. In the experiments, we tested three values of ratio $r$ that may fit a wide range of real-world situations available in actual social networks: a conservative $r = 0.2$ (all users share more neutral than misinformative items), an unbiased $r = 0.5$, and an extreme $r = 0.8$. Table 2 shows statistics about the generated datasets according to these ratios. We note the especially low density values, in particular compared against standard datasets in recommendation, whose density ranges between 4 and 6% [65]. We leave as future work to experiment with users of different ratios co-existing in the same simulation of the system.

Table 2: Statistics from the obtained datasets according to the methodology presented in Section 5.1. *Density* accounts for the number of cells with information in the user-item matrix, that is, $R/(U \cdot I)$, considering $R$ the number of interactions and $U$ and $I$ the number of users and items.

| Ratio | Users | Items | Interactions | Density (%) |
|---|---|---|---|---|
| $\emptyset$ | 2,921 | 1,014,004 | 1,116,658 | 0.038 |
| 0.2 | 2,919 | 28,378 | 33,065 | 0.040 |
| 0.5 | 2,921 | 5,761 | 10,084 | 0.060 |
| 0.8 | 1,999 | 914 | 3,909 | 0.214 |

Table 3: Parameters of evaluated recommendation algorithms. Values in bold denote the *typical* parameterisation that will be referenced later. For MF, $k$ denotes the number of factors, $\lambda$ controls the overfitting, and $n$ is the number of iterations. For UB and IB, $k$ denotes the number of neighbours, *sim* is the similarity, and $q$ is the exponent of similarity value.

| Rec | Parameters |
|---|---|
| MF | $k$ = {20,**50**,100}, $\lambda$ = {**0.1**,0.01}, $n$={**20**,100} |
| IB | $k$ = {10,**50**,100}, sim = {jac, cos, **pearson**}, $q$={**1**,2,3} |
| UB | $k$ = {10,**50**,100}, sim = {jac, cos, **pearson**}, $q$={**1**,2,3} |

*5.2. Recommendation Algorithms*

As discussed throughout the paper, the recommendation algorithms are at the core of the so-called feedback loop. Here, we study how different families of algorithms may amplify misinformation under different initial constraints, as described in the previous section. For this, we focus on the most common algorithmic techniques to produce recommendations, namely collaborative filtering approaches [66].

These techniques have the main advantage that do not depend on user or item metadata or attributes, since they only require the user-item interactions to model the user preferences and, based on that, to produce suggestions. Because of this, they are widely used in several domains, ranging from movie or music recommendation to the travel domain [67, 68, 69]. However, they are well-known to suffer from popularity bias, or the *rich gets richer* effect [15, 14]. Therefore, they are good candidates to analyse if popularity bias translates into a potential misinformation spreading, or under which conditions this is more likely to occur.

To properly understand this behaviour, we selected three classical methods that are well-known and widely used in both academia and industry:

- A **matrix factorisation algorithm (MF)** [70] that uses Alternate Least Squares in its minimisation formula. This method learns latent factors for users and items, and try to reconstruct the original user-item interaction matrix by minimising the distance between the original and reconstructed matrices.

- A **nearest neighbour algorithm based on users (UB)**, which exploits similar-minded users from the community [31] to produce the recommendations [71]. We use a non-normalised version as follows since it has shown better ranking performance [72]:

$$s(u,i) = \sum_{v \in N(u;k)} s(v,i) w(u,v)^q \tag{1}$$

  where $N(u;k)$ denotes the $k$ closest users (neighbours) in terms of similarity to user $u$, $w(u,v)$ is a similarity function, and $q$ is a weight to emphasise the value of such similarity.

- A **nearest neighbour algorithm based on items (IB)**, which, similarly to UB, generates recommendations according to a neighbourhood, but in this case by exploiting similar items to those previously interacted by the user

---

[31] We note that the term *neighbourhood* is used as in classical recommendation, to denote users selected according to the similarity, and it has no relation with the structure of the Twitter network.

[71]. We also use a non-normalised version, whose formulation is the following:

$$s(u,i) = \sum_{j \in N(i;k)} s(u,j)w(i,j)^q \qquad (2)$$

Besides, we also included a random (Rnd) and a most-popular (Pop) baselines that provide non-personalised recommendations while controlling biases in the data: whereas Rnd will produce completely unbiased suggestions, Pop will be guided purely by the items with more interactions (independently, in principle, of their misinformative nature).

Since there is no training-test split in our experiments, we cannot optimise any accuracy metric to select the hyperparameters. To overcome this limitation, we experimented with some typical parameterisations of these approaches, together with other variations, all of them presented in Table 3.

*5.3. Evaluation Metrics*

The goal of our work is to determine whether and to what extent the recommendation algorithms help to spread misinformative items. We thus disentangle the efficiency (i.e., accuracy) dimension typically associated to the performance evaluation of these algorithms from the actual measurement of the misinformation spread. As a consequence, we do not divide our data into training and test sets, which, considering the small number of users and interactions (compared with typical datasets in the recommender systems community) in some of our simulated scenarios (see Section 5.1), allows us to make the most out of this information, avoiding to holdout any data point for validation or testing.

With this goal in mind, we propose the following three evaluation metrics that measure how many misinformative items are present in the recommendation lists provided by the RAs, either in an absolute way (count), compared against the user prior distribution (ratio difference), and from a global perspective (Gini):

- **Misinformation Count (MC)** measures how many of the recommended items to a user are misinformation. Its value is normalised by a cutoff, which corresponds to the ranking size, so that comparable measurements could be produced at different sizes. The higher the MC value, the more misinformation included in the recommendations; its range is in [0,1].

- **Misinformation Ratio Difference (MRD)** computes how much the ratio of misinformation has changed in a user basis with respect to what is observed in training. In particular, we calculate the ratio of misinformation in training for each user (let us call it $m^u_t$), and compare it with the observed ratio of misinformation in the recommendation list ($m^u_r$). Then, this metric is the average of the differences $m^u_t - m^u_r$. Hence, the larger the MRD value in absolute terms, the higher the change with respect to training, whereas its sign indicates the direction of such change: a positive value one would show that the ratio is larger in training. The range of this metric is [−1,1].

**Misinformation Gini (MG)** measures the dispersion over a distribution, as it is done to account for diversity in recommendation [73]. In our case, we compute the distribution over the misinformative items by considering the number of times each item was recommended, and add another item that represents the rest of the items in the collection, i.e., the neutral or informative ones. In this way, when the distribution is uniform (all the misinformative items have been recommended a similar number of times), MG would produce a higher value than when the above distribution is highly skewed. Note that, in contrast to the other two metrics, this one is not computed in a user basis, but for the entire set of recommendations. Its range is [0,1].

Table 4: Misinformation metrics for the typical configurations of CF recommender systems at different ratios of misinformation present in the user profiles. MC, MG, and MRD denote count, Gini, and ratio difference of misinformation, as presented in Section 5.3.

| Ratio | Rec | MC@5 | MC@10 | MC@20 | MRD@5 | MRD@10 | MRD@20 | MG@5 | MG@10 | MG@20 |
|---|---|---|---|---|---|---|---|---|---|---|
| ∅ | Rnd | 0.001 | 0.001 | 0.001 | 0.031 | 0.031 | 0.031 | 0.000 | 0.000 | 0.000 |
| ∅ | Pop | 0.000 | 0.002 | 0.098 | 0.032 | 0.030 | −0.065 | 0.000 | 0.000 | 0.000 |
| ∅ | MF | 0.053 | 0.040 | 0.032 | −0.021 | −0.007 | 0.000 | 0.001 | 0.001 | 0.001 |
| ∅ | IB | 0.018 | 0.013 | 0.008 | 0.014 | 0.020 | 0.024 | 0.000 | 0.000 | 0.000 |
| ∅ | UB | 0.068 | 0.054 | 0.044 | −0.036 | −0.022 | −0.012 | 0.006 | 0.006 | 0.006 |
| 0.2 | Rnd | 0.027 | 0.026 | 0.026 | 0.109 | 0.110 | 0.110 | 0.009 | 0.012 | 0.016 |
| 0.2 | Pop | 1.000 | 1.000 | 1.000 | −0.864 | −0.864 | −0.864 | 0.006 | 0.012 | 0.026 |
| 0.2 | MF | 0.995 | 0.984 | 0.919 | −0.859 | −0.848 | −0.783 | 0.189 | 0.222 | 0.237 |
| 0.2 | IB | 0.091 | 0.063 | 0.047 | 0.049 | 0.077 | 0.093 | 0.005 | 0.004 | 0.005 |
| 0.2 | UB | 0.327 | 0.213 | 0.131 | −0.188 | −0.073 | 0.009 | 0.054 | 0.041 | 0.028 |
| 0.5 | Rnd | 0.133 | 0.131 | 0.131 | 0.367 | 0.369 | 0.369 | 0.088 | 0.098 | 0.106 |
| 0.5 | Pop | 1.000 | 1.000 | 1.000 | −0.500 | −0.500 | −0.500 | 0.006 | 0.012 | 0.026 |
| 0.5 | MF | 1.000 | 0.998 | 0.970 | −0.500 | −0.498 | −0.470 | 0.203 | 0.246 | 0.266 |
| 0.5 | IB | 0.132 | 0.112 | 0.100 | 0.368 | 0.387 | 0.396 | 0.012 | 0.017 | 0.027 |
| 0.5 | UB | 0.340 | 0.235 | 0.217 | 0.160 | 0.264 | 0.279 | 0.059 | 0.051 | 0.064 |
| 0.8 | Rnd | 0.759 | 0.757 | 0.755 | 0.226 | 0.228 | 0.230 | 0.599 | 0.631 | 0.650 |
| 0.8 | Pop | 1.000 | 1.000 | 1.000 | −0.015 | −0.015 | −0.015 | 0.006 | 0.012 | 0.026 |
| 0.8 | MF | 1.000 | 0.995 | 0.969 | −0.015 | −0.010 | 0.016 | 0.221 | 0.265 | 0.300 |
| 0.8 | IB | 0.667 | 0.627 | 0.515 | 0.252 | 0.194 | 0.154 | 0.280 | 0.348 | 0.375 |
| 0.8 | UB | 0.897 | 0.766 | 0.586 | 0.022 | 0.054 | 0.082 | 0.286 | 0.334 | 0.356 |

## 6. Results

In this section, we report and discuss the results of the evaluation metrics presented in Section 5.3 on different scenarios where the ratio of misinformation has been configured as explained in Section 5.1. First, in Section 6.1, we explore the effect on misinformation spread of the most common instantiations of the analysed recommendation algorithms; then, in Section 6.2, we perform a sensitivity analysis on the effect of the parameters of such algorithms for misinformation spreading. Some limitations of these experiments are discussed in Section 8, after guidelines for adaptation are proposed in Section 7.

Table 5: Misinformation count measured at cutoff 10 aggregating the results according to the amount of information used by each algorithm: number of factors in MF and neighbours in IB and UB.

| Rec | Info | ∅ | 0.2 | 0.5 | 0.8 |
|---|---|---|---|---|---|
| MF | High | 0.077 | 0.907 | 0.953 | 0.959 |
| MF | Med | 0.062 | 0.988 | 0.999 | 0.995 |
| MF | Low | 0.015 | 0.997 | 1.000 | 1.000 |
| IB | High | 0.016 | 0.121 | 0.213 | 0.668 |
| IB | Med | 0.016 | 0.122 | 0.211 | 0.667 |
| IB | Low | 0.017 | 0.136 | 0.213 | 0.662 |
| UB | High | 0.057 | 0.242 | 0.272 | 0.772 |
| UB | Med | 0.055 | 0.213 | 0.235 | 0.766 |
| UB | Low | 0.038 | 0.048 | 0.072 | 0.469 |

Table 6: Misinformation count measured at cutoff 10 aggregating the results according to the additional information exploited by each algorithm: number of iterations in MF and similarity weight $q$ in IB and UB.

| Rec | Info | ∅ | 0.2 | 0.5 | 0.8 |
|---|---|---|---|---|---|
| MF | High | 0.077 | 0.997 | 1.000 | 1.000 |
| MF | Low | 0.074 | 0.997 | 1.000 | 1.000 |
| IB | q=1 | 0.016 | 0.136 | 0.213 | 0.668 |
| IB | q=2 | 0.017 | 0.136 | 0.213 | 0.665 |
| IB | q=3 | 0.017 | 0.136 | 0.213 | 0.664 |
| UB | q=1 | 0.057 | 0.242 | 0.272 | 0.772 |
| UB | q=2 | 0.042 | 0.098 | 0.123 | 0.764 |
| UB | q=3 | 0.037 | 0.052 | 0.088 | 0.761 |

*6.1. Misinformation Spread of Recommendation Algorithms*

Table 4 shows our proposed misinformation evaluation metrics (count, ratio difference, and Gini) when testing different configurations of the misinformation ratio to create the user profiles. For these experiments, we tested the RAs presented in Section 5.2 using their most typical configuration of parameters (as shown in Table 3), which have demonstrated to be very effective on different domains.

Based on this, our first observation is that, except when ratio is ∅ (that is, when no control on the amount of misinformation interacted by the users is imposed), the popularity-based Pop algorithm is the most effective method in spreading misinformation, both in terms of MC (where all the items presented to the user are labelled as misinformative) and MRD (where the method produces the most negative differences with respect to the misinformation ratio in training, meaning that it increases such ratio for all users consistently). The reason for this might be obvious: once we force all users to have at least 20% of their items to be misinformative, it is more likely that the most popular items in the system are, at the same time, misinformative.

Interestingly, these results evidence that our simulations with a positive misinformation ratio produce situations where a small number of misinformative items get popular very quick. This is indeed quite realistic, as it often occurs in social media where fake news or other dubious pieces of information are spread rapidly. However, and according to our results, such spread can be slowed down with an appropriate use of recommendation algorithms, as we shall see next.

Besides the random Rnd recommender, which usually includes the lowest number of misinformative items in its suggestions due to its complete disregard of the interactions between users and items, we observe that the methods based on neighbours (UB and IB) spread less misinformation. We should note that the Rnd recommender is actually reflecting the distribution of the population, hence, in these cases, *most of the items are not misinformative*, which is true by design except when $r = 0.8$. The methods based on neighbours, which are expected to produce more relevant personalised recommendations than the Rnd recommender, are able to keep the spread of misinformation between 10 and 30%, as long as the original ratio of misinformation in the user profiles is not too high, that is, for $r = 0.2, 0.5$. This is in contrast with the Pop and the MF recommenders. The latter algorithm basically *follows* the Pop method, in the sense that in those configurations where the ratio is positive, it produces results very close to those obtained for Pop. This can be attributed to a strong popularity bias evidenced by this and other algorithms in the area, a well-studied problem by the community [15]: in general, good results are obtained when producing popular but slightly personal results for each user, even though the utility of such recommendations is very limited, and hence, a tradeoff between novel, diverse, and popular items is demanded by the users [73]. In this work, we can add another negative consequence of this behaviour: a larger presence of misinformative items in the recommendations.

Moreover, by analysing the misinformation Gini metric, we can better understand the differences between these two approaches. Recall that MG measures how uniform the distribution of misinformative items is, from a global perspective. Hence, since MF obtains higher values than Pop, these results show that Pop is always recommending the same (limited) set of the misinformative items. MF, in contrast, is recommends a wider range of items, even though most of them turn out to also be misinformative. In this sense, we could infer that the spread of misinformation is different for these algorithms: Pop is very aggressive on suggesting the same items over and over again, as if it was a bot or a viral account in the system; on the other hand, MF distributes more evenly the misinformative items across the population, hence spreading out a larger number of distinct misinformative items.

Nonetheless, most of these observations dramatically change when there are no constraints on the misinformation ratio. In our Table 4, when ratio is ∅ we observe that the popularity-based recommender does not longer spread misinformative items in the same way. This is attributed to the items being less common among the entire population and, hence, not popular enough to be recommended. However, MF and, surprisingly, UB seem to be very effective in recommending a significantly large number of misinformative items (especially, if we compare against the random recommender) also in this situation. Our initial conclusion, hence, is that the MF algorithm, independently of the starting scenario, will increase the presence of misinformative items due to its recommendations, which is exacerbated in the long term if we consider recommendation algorithms as part of the feedback loop. Neighbour-based methods, and in particular, IB seem to be safer in this respect, since they tend to control the spread under some reasonable limits. In the next section, we continue our analysis exploring how sensitive the recommenders are to spreading misinformation when different values of their model parameters are used.

*6.2. Effect of Recommendation Parameters in Misinformation Spread*

Table 5 shows a complementary analysis of the results presented before but only for the metric MC@10, as it is the easiest to interpret. In this table, we aggregate the performance obtained by the RAs focusing on the amount of information exploited by each algorithm. This translates into the number of factors for MF and number of neighbours for UB and IB. This means that a row where High appears in the Info column aggregates the values of all the recommenders of the same type whose number of factors or neighbours are above some predefined threshold. In particular, for this analysis and considering the parameters shown in Table 3, we consider 100 factors or neighbours as High, 50 factors or neighbours as Med, and 20 factors or 10 neighbours as Low.

According to these results, we observe that under controlled conditions (i.e., a positive misinformation ratio), the number of factors or neighbours do not have a strong effect in changing the spread of misinformation for MF or IB. For UB, in contrast, a low number of neighbours drastically reduces the number of misinformative items being recommended. This observation might be linked to the previous discussion on popularity bias: as investigated in the Recommender Systems area [74], UB with large neighbourhoods tends to be closer to popularity, in this case, a lower value allows to recommend less popular items which, in the controlled conditions, are more likely to not be misinformative (by design, as discussed in the previous section).

This effect, interestingly, is also observed when no constraints on misinformation ratios are imposed. Therefore, we conclude that a low number of neighbours in UB could be helpful to stop the spread of misinformative items under all the conditions we have tested in this work. Additionally, we also observe in Table 5 that a lower number of factors in MF limits the number of misinformative items recommended by this algorithm. This behaviour, however, is inconsistent in the rest of simulated conditions, where a low number of factors ensures that almost all the recommended items will be misinformative.

Table 6 shows a summary of results similar to that of Table 5, but considering other parameters of the algorithms. In this case, for MF we analyse the number of iterations (High is used for 100, Low for 20) and for UB and IB the similarity weight $q$ used in Equations 1 and 2 to determine how much each neighbour influences the final prediction.[32] From the results, similarly to the analysis shown in Table 5, we observe that neither MF nor IB seem to be affected by the above parameters in their abilities to increase the spread of misinformation. However, a large $q$ in UB consistently reduces the amount of misinformative items recommended by this algorithm.

**7. Adaptation Guidelines**

In this section, we propose a series of adaptation guidelines for RAs based on our analysis, as well as on the review of the research literature. Inspired by the field of context-based RAs [75], we argue that RAs can be adapted at three different points: (i) pre recommendation, (ii) within the model, and (iii) post recommendation.

One key conclusion of our analysis is that, if we want to palliate the misinformation amplification effect of RAs, we need to reduce the popularity effect. This can be achieved by selecting RAs based on neighbours (like UB and IB), and by reducing the number of neighbours used within the models. The more neighbours the RA uses, the more it resembles popularity. Our recommendation for adaptation is consequently to reduce the number of neighbours. For example, one could think about RAs that cluster the users' social contacts in different subgroups according to similarity (where the first cluster contains the most similar social contacts, and the last cluster the more dissimilar contacts). The algorithm could then recommend a ratio of items from each of those clusters. That will mean that (i) we will be reducing the popularity effect on one hand, since the neighbours are smaller, and (ii) we could be introducing some diversity in the recommendations provided (since we could even recommend some items from the user's most dissimilar social contacts, taking into consideration not introducing a large cognitive dissonance).

Tools that allow for the detection of malicious actors (bots, sockpupets) could also help us to adapt RAs by doing pre or post adaptations. For example, one may use tools like Botometer or misinfo.me to discard all user accounts that resemble bots, or that frequently spread misinformation, and do not take into account any of their content for recommendations. On the other hand, the adaptation could be done a posteriori, by re-ranking recommended items based on the "reliability" of the account from where those posts originated. This relates to the notion of trust in RAs, although in a slightly different manner, since trust is normally considered among two users [76], while here we are referring to the reliability of accounts. The same could be thought about content. Content that ranks low on credibility could be either discarded (pre-adaptation) or re-ranked (post-adaptation). These elements could also be addressed at the model level, incorporating scores of reliability and credibility of users and content as part of the user and item profiles. This would enable more sensible adaptations, like diluting the weights of potentially misinforming users and items over cycles of recommendation.

---

[32] As noted in [72], a higher value of $q$ will make smaller similarities drop to 0, while higher ones will be (relatively) emphasised.

Further adaptations could be done on the modelling of users and items by incorporating some of the elements discussed in Sections 2.1.1 and 2.1.2. Previous studies have shown how personality, values, emotions and vulnerability of users affect their likelihood to propagate misinformation. Considering these aspects when profiling users and items, could help RAs to be more selective on their recommendations.

When thinking about adaptation of RAs it is also important to consider those strategies that have been proven effective when correcting misperceptions (see Section 2.2). RAs could be adapted to promote corrective information without introducing a high-degree of cognitive dissonance (e.g., by providing corrections that are "observational" –over topics where the user is not emotionally invested) [54] or providing corrections from users (social connections) that are similar, revealing the similarities of the opposing group [49].

## 8. Discussion and Limitations

The goal of the presented study has been to analyse the impact of RAs on the spread of misinformation in social networks, and particularly Twitter. As we have seen by the analysis of the literature (see Section 2), this constitutes, to the best of our knowledge, a novel and exciting research strand. It is, however, one where multiple challenges emerge and need to be confronted (see Section 3).

One of these challenges is the generation of datasets. As we have reported in Sections 4 and 5, the generated dataset, although the first of its kind, suffers from several biases and limitations. Users have been selected based on the Coronavirus Facts Alliance Dataset. Although this dataset covers misinformation (identified by fact-checkers) from 74 countries, it is specific to COVID-19. Users who never spread COVID-19 misinformation are not covered in our data. Users in our dataset have very low numbers of misinforming items in their timeline, which may not be entirely representative of a social network. One may expect to see some users spreading a lot of misinformation, while many may spread none [64]. The creation of synthetic datasets could be a potential solution for future work, even though real effects would need to be measured via user studies, which opens up new challenges in terms of privacy and ethics. Twitter also announced, two days before the submission deadline of this article,[33] that current API restrictions will be removed for academic research, allowing the collection of historical data. This is a great initiative that will enable the creation of more comprehensive, complete and unbiased datasets.

Our study has focused on the analysis of RAs based on Collaborative Filtering techniques. Analysing content-based and Hybrid methods requires to capture RAs dealing with natural language and its inherent subtleties (negation, sarcasm, etc.). An in-depth algorithmic survey is therefore required to better understand the impact of these techniques in the recommendation of misinformation. This includes classical and hybrid collaborative algorithms [77] and more recent methods aimed at understanding the natural language by, for instance, using Neural Networks [78].

Our results have shown that it is possible to limit the inherent spread of misinformation derived from RAs by configuring these techniques in different ways. However, it should be emphasised that no tradeoff with respect to the potential loss (or gain) on accuracy derived by such changes was measured. This is important since the Recommender Systems community has shown that several beyond-accuracy dimensions compete between each other and against accuracy when designing the perfect user experience, and it is extremely difficult to find an algorithm that is optimal for more than one dimension at the same time [79, 73]. Nonetheless, in this work we wanted to focus strictly on the spread of misinformation, so we decided to isolate the problem and study it independently. Moreover, considering the difficulty on collecting the data for our study, as presented in Section 4, not needing to separate the data into training and test (for classical evaluation of the recommendation algorithms) allowed us to devote more data to the purpose of the study. We hope to conduct analyses with accuracy measurements in the future by enriching our current dataset and/or generating synthetic ones.

It is also important to highlight that accuracy and metrics that target user satisfaction, may not be the most effective ones when aiming to reduce the impact and spread of misinformation. Algorithms promoting a certain degree of cognitive dissonance, as suggested by existing literature on correcting misperceptions (see Section 2.2), and metrics that focus on computing a balanced degree of user satisfaction and discomfort, may be more suitable to combat misperceptions.

Moreover, it is worth noting that the results presented so far only involve one cycle in the feedback loop introduced at the beginning of the paper. Because of this, some of the results that we obtained, such as reducing the spread to a range of 10-30%, instead of 90%, may not be enough if the users engage in several cycles of receiving recommendations and interacting with them. In fact, we have simulated another cycle of recommendation in Table 7, where we contrast how the misinformation evolves starting from the same data (user profiles built with a misinformation ratio of 0.2) and running two recommenders (MF and UB) after we assume that all users accept their top-3 recommendations, either those produced by MF or UB. As we observe, the number of misinformative items in the recommendations would increase steadily at each recommendation cycle, although this speed is much lower for UB than for MF.

---

[33] https://blog.twitter.com/developer/en_us/topics/tools/2021/enabling-thefuture-of-academic-research-with-the-twitter-api.html

Table 7: Evolution of MC@10 depending on which recommender is used to present items to users. For $t = 2$, we simulate that all users accept their top-3 recommendations, train the recommendation algorithms again, and measure the misinformation of the items returned by each method.

| Rec. cycle | HKV | UB |
| --- | --- | --- |
| t=0 | 0.200 | 0.200 |
| t=1 | 0.984 | 0.213 |
| t=2 (after UB) | 0.658 | 0.355 |
| t=2 (after HKV) | 0.988 | 0.762 |

An important point to make while presenting this research is the need of ethical guidelines [80]. We need to be very careful when adapting existing algorithms to ensure that we do not introduce damaging effects. For example, algorithmic adaptations that may reduce the recommendation of misinformation, but that tend to promote misinformation of a more harmful nature should not be considered successful. This research requires to navigate the careful tension between privacy, security, economic interests, censorship and cultural differences, and requires to be addressed from multiple disciplines that can assess not only the technological aspect, but also the individual and the social one. As discussed, there is ample room for investigation in the proposed work, opening a novel, exciting and interdisciplinary line of research.

## 9. Conclusions

Recommendation algorithms have been pointed out as one of the major culprits of misinformation spreading in the digital sphere. This paper investigates the effect of RAs on the spread of misinformation in social networks, particularly in Twitter. Our results indicate that RAs that suffer from popularity bias, as well as algorithmic settings that end up mimicking popularity, are more likely to spread misinformation. Adaptations that try to reduce the popularity effect, as well as the incorporation of pre, model, and post adaptations based on the existing literature of misinformation management are also proposed. A dataset has also been generated and released to the scientific community to stimulate discussions and further work.

## Acknowledgements

This work has been co-funded by H2020 Co-Inform (ID:770302) and HERoS (ID:101003606) projects and the Spanish Ministry of Science and Innovation (project reference: PID2019-108965GB-I00).